\documentclass[amsmath,amssymb,reprint,aps,prl,showpacs,superscriptaddress,floatfix]{revtex4-1}
\usepackage{graphicx}
\usepackage{color}
\usepackage{epstopdf}
\usepackage{braket}
\usepackage{siunitx}
\usepackage[normalem]{ulem}

\makeatletter

\begin{document}

\title{Self-assembly of tetracyanonaphtho-quinodimethane (TNAP) based metal-organic networks on Pb(111): Structural, electronic, and magnetic properties}
\author{Gelavizh Ahmadi}
\author{Katharina J. Franke}

\affiliation{Fachbereich Physik, Freie Universit\"at Berlin,
                 Arnimallee 14, 14195 Berlin, Germany.}
\date{\today}

\begin{abstract}

We use scanning tunneling microscopy and spectroscopy to investigate structural and electronic properties of tetracyanonaphtho-quinodimethane (TNAP) based metal-organic networks on a superconducting Pb(111) surface. At low temperatures, the TNAP molecules form densely packed islands. When deposited at room temperature, Pb adatoms are incorporated into fourfold bonding nodes with the TNAP molecules leading to long-range ordered porous structures. Co-deposition of NaCl with TNAP yields a Na source for an ionically bonded Na-TNAP structure. Fourfold bonding motifs are also created by Fe atoms with the cyano terminations of TNAP. However, the structures are irregular and do not sustain the formation of long-range ordered networks. Some Fe centers with molecules surrounded in a local C2 symmetry exhibit Shiba states as a fingerprint of a magnetic interaction with the superconducting surface.

\end{abstract}

\maketitle 

\section{Introduction}
The adsorption of molecular layers is one of the key strategies to confer a desired functionality to a surface. The construction of long-range ordered structures relies on self-assembling processes, which are driven by attractive forces between the adsorbates. Interactions of the adsorbates with the substrate further influence the resulting structure and electronic properties. Enormous amounts of examples of self-assembled homo-molecular~\cite{BarthAnnRevPhysChem07, CicoiraTopCurrChem08, LiangCoordChemRev}, hetero-molecular~\cite{LiangCoordChemRev,OteroAdvMat11, ZhangNanoscale15}, as well as metal-organic structures~\cite{LiangCoordChemRev, BarthSurfSci07, LinTopCurrChem09} have been reported on noble metal surfaces. Their envisioned functions demand the formation of templates for adsorption cavities~\cite{TheobaldNature03, WintjesAngew07, PivettaPRL13, PalmaNanoLett14}, the tailoring the electronic structure for tuning the dispersion of surface bands~\cite{LakunzaPRL08, LoboScience09, KlappenbergerPRL11}, the stabilization of magnetic properties~\cite{GambardellaNatMat08, UmbachPRL12, AbdurakhmanovaPRL13} for information storage. 
New perspectives arise, if the substrate was not a simple metal, but a superconductor. This is particularly interesting for the design of magnetic networks on a superconductor. A single magnetic adsorbate can serve as a scattering potential for the quasiparticles of the superconductor, i.e., the Cooper pairs. This gives rise to so-called Yu-Shiba-Rusinov states~\cite{Yu65, Shiba68, Rusinov69}, which appear as transport resonances inside the superconducting energy gap~\cite{YazdaniScience97, JiPRL08, FrankeScience11}. Recently, it has been pointed out theoretically that a lattice of such magnetic scattering centers leads to intriguing topological properties~\cite{RoentynenPRL15}. It has also been emphasized that, in particular, metal-organic structures on superconducting Pb could possibly lead to topological conditions, which bear Majorana states~\cite{LuArxiv15}.

Here, we explore the self-assembly of metal-organic structures on the superconducting Pb(111) surface.  As a promising molecular linker for metallic atoms, we chose 11, 11, 12, 12-tetracyanonaphtho-2, 6-quinodimethane (TNAP). This molecule is well suited as a building block for metal-organic networks with alkali atoms as well as transition metal atoms~\cite{KanaiJApplPhys09}. On the one hand, the capability to form self-assembled networks with alkali atoms can be ascribed to its large electron affinity ($E_A=4.7$~eV), which tends to withdraw the valence electron from the alkali atom to form an ionic bond. On the other hand, the cyano terminations with their lone-pair electrons are prone to form coordination bonds to transition metal atoms~\cite{LiangCoordChemRev, BarthSurfSci07, LinTopCurrChem09}. 

We first characterize the structural and electronic properties of pure TNAP on the Pb(111) surface. The mobility and electronic flexibility to sustain different bonding symmetries of Pb atoms leads to a Pb-TNAP network formed at room temperature, where Pb forms the bonding node between the electrophilic cyano terminations. We further show that Na atoms provided by co-deposited sodium chloride (NaCl) islands lead to long-range ordered Na-TNAP structures, which are probably stabilized by an ionic bonding nature. Co-deposited Fe atoms are also incorporated into bonding nodes with the cyano groups. However, we could not observe long-range ordered structures on Pb(111). We show that the local symmetry of the Fe coordination drastically affects the magnetic properties. Only some Fe atoms surrounded by TNAP molecules in a local C2 symmetry show Shiba resonances as a fingerprint of magnetic interaction with the superconductor.

\section{Experimental details}
All experiments were carried out under ultra high vacuum conditions with the base pressure of $<10^{-9}$~mbar. An atomically clean Pb(111) surface was prepared by repeated cycles of Neon ion sputtering and subsequent annealing to 530~K. TNAP and NaCl were deposited from a Knudsen cell at 450~K and 810~K, respectively. Iron was sublimed from an ultra pure rod (99.99\% purity) by electron bombardment. The sample was kept at the indicated temperature for each experiment and eventually post-annealed.
The sample was then cooled and transferred into a custom-build scanning tunneling microscope, operated at 4.5~K. Tunneling spectra were acquired with a lock-in amplifier at a modulation frequency of 833~Hz. The modulation voltages and feedback conditions are indicated in the figure captions. The STM tip was covered by a thick layer of Pb by controlled indentation into the surface while applying 100~V. This procedure was repeated until the tip exhibited bulk-like Pb properties, in particular a superconducting energy gap around the Fermi level~\cite{FrankeScience11}.

\section{Results and discussion}

\begin{figure}
\includegraphics[width=8cm,clip=]{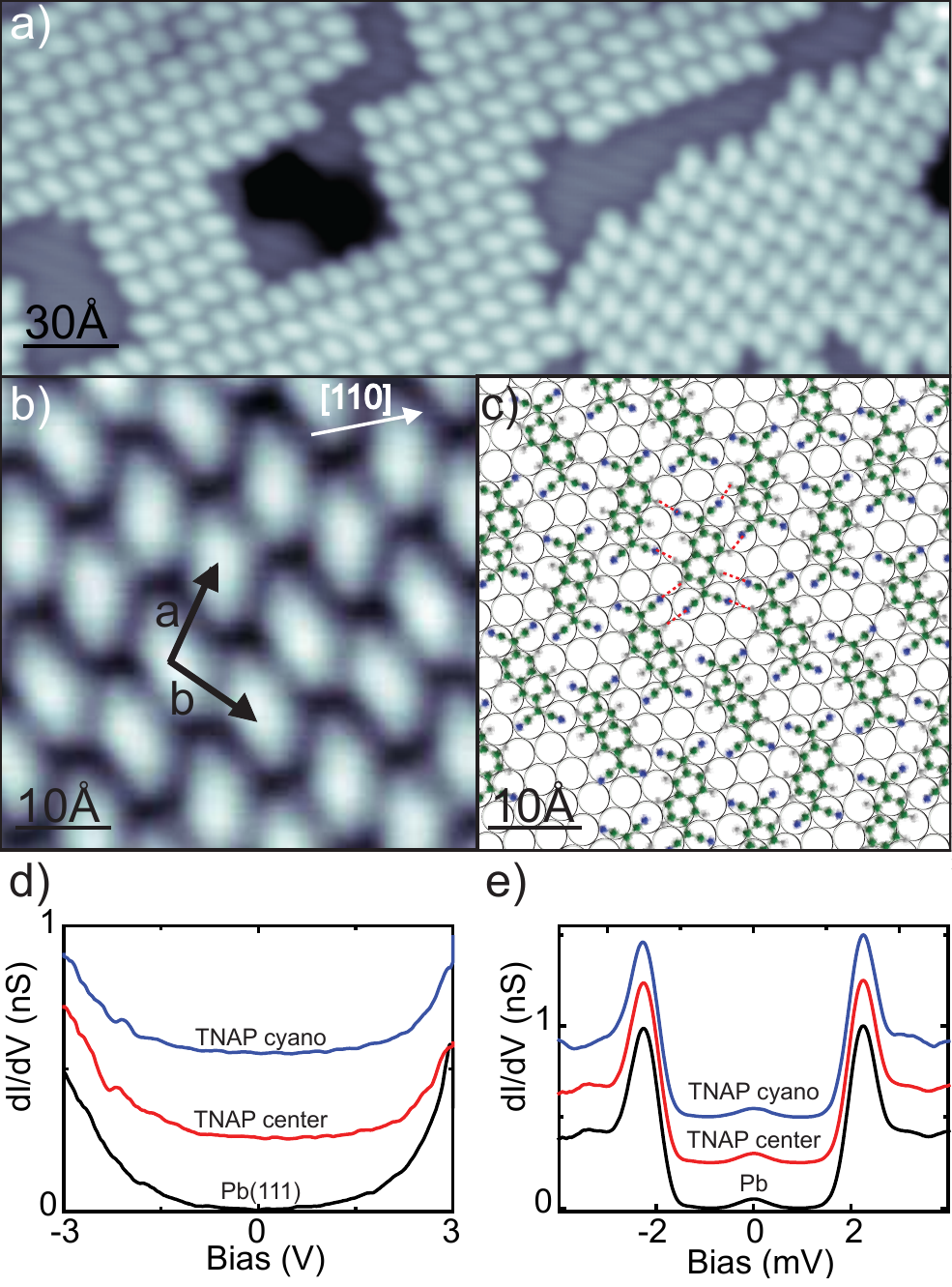}
\caption{a) STM topographic image of TNAP islands on Pb(111)($I=0.2$~nA, $U=290$~mV) b) STM image of a TNAP island with the rhombic unit cell indicated by arrows ($I=1.1$~nA, $U=-30$~mV). c) Structure model of a TNAP island on top of Pb(111). d) Constant-height $dI/dV$-spectra acquired over the different  sites of the TNAP molecules and on the clean Pb(111) surface (feedback opened at $I=0.2$~nA, $U=3$~V  with the modulation of $V_\mathrm{rms}=5$~mV). e) Constant-height $dI/dV$-spectra around the Fermi level over the different  sites of the TNAP molecules and on the clean Pb(111) (feedback opened at $I=0.23$~nA, $U=10$~mV  with the modulation of $V_\mathrm{rms}=0.1$~mV). The spectra in (d, e) are offset for clarity. The STM images were prepared using the software WSxM~\cite{wsxm}.}
\label{fig1} 
\end{figure}

\subsection{TNAP on Pb(111)}

Deposition of a submonolayer of  TNAP onto Pb(111) at 160~K leads to the formation of well-ordered molecular islands as shown in Fig.~\ref{fig1}(a). The islands exhibit a preferential orientation with the molecular rows aligned along the [110] direction of the Pb lattice~\cite{note1}.  
A close-up view into the islands reveals the size and shape of the molecules [Fig.~\ref{fig1}(b)]. The molecules are separated by $a=b=10\pm1~{\textrm{\AA}}$, where $a$ and $b$ enclose an angle of $\alpha=98\pm2^{\circ}$. This allows for a flat-lying adsorption on the substrate, which is typical for organic molecules with an extended $\pi$ electron system along a flat backbone \cite{BarthAnnRevPhysChem07, OteroAdvMat11}. The corresponding structural model in Fig.~\ref{fig1}(c) is very similar to TNAP islands on gold~\cite{KanaiJApplPhys09, UmbachJPCM12, FiedlerJPCC14}. Intermolecular electrostatic forces from the electronegative N atoms to the H atoms of the quinodimethane unit of the neighboring molecule probably stabilize the densely packed islands on both surfaces. Since the difference in the two enantiomers is minute, both of them are statistically incorporated in the islands, as can be recognized from the slighly different orientation of the molecules, similar to the case on Au(111)~\cite{FiedlerJPCC14}.
Whereas no commensuration of the islands was found on Au(111) \cite{UmbachJPCM12, FiedlerJPCC14}, the molecules are in registry with every fourth Pb atom in the [110] direction. This commensuration of the two lattices probably contributes to the stabilization of the molecular islands.

TNAP is a model molecular charge acceptor with an electron affinity of $E_\mathrm{A}=4.7$~eV~\cite{KanaiApplPhysA09}. When brought into contact with a Pb surface, which exhibits a workfunction of $\Phi_\mathrm{Pb}=3.8$~eV~\cite{YuPRB04}, one may expect a significant amount of charge transfer~\cite{BraunAdvMat09}. 
To elucidate on the molecular energy level alignment and possible traces of charge transfer, we record scanning tunneling spectra (STS) on the cyano groups and the molecular center. None of them exhibits obvious resonances in the energy range of $\pm 3$~eV [Fig.\ref{fig1}(d)]. In search for a sign of charge transfer we also examine the region of the superconducting energy gap. Pb is a type I Bardeen-Cooper-Schrieffer (BCS) superconductor with a critical temperature $T_\mathrm{c}=7.2$~K. At our measurement temperature of 4.5~K the superconducting state is characterized by an energy gap of $2\Delta=2.2$~meV.  
Differential conductance spectra recorded with a superconducting tip on a superconducting substrate reflect the BCS-like resonances as two sharp peaks at an energy of $\pm(\Delta_\mathrm{tip}+\Delta_\mathrm{sample})$ [see Fig.~\ref{fig1}(e)]. The small peak at the Fermi level is due to tunneling of thermally excited quasiparticles \cite{FrankeScience11}. 
In the case of a single electron being localized in a TNAP orbital one may expect that its exchange interaction with the superconducting substrate gives rise to Shiba states, which are an expression of an anti-ferromagnetic coupling of an unpaired electron spin with the Cooper pairs of the superconductor~\cite{Yu65, Shiba68, Rusinov69}. We do not find a hint of such resonances inside the superconducting energy gap [Fig.~\ref{fig1}(e)]. 

The absence of any such spectroscopic signatures could be indicative of a large molecular energy gap around the Fermi level. However, this is very unlikely, because the free molecule has an energy gap of $E_\mathrm{g}=1.95$~eV~\cite{KanaiJApplPhys09}, which is typically reduced on a metal surface~\cite{TorrenteJPCM08}. Instead, we suggest a completely different scenario, which can be derived from a simple model of energy level alignments considering $\Phi_\mathrm{Pb}<E_\mathrm{A,~TNAP}$. Due to the smaller workfunction, electron density is transferred from the substrate to the TNAP molecules~\cite{IshiiAdvMat99}. The amount of charge transfer is determined by the electrostatic potential and extension of wavefunctions at the interface. Presumably, this leads to a non-integer charge state with considerable hybridization of molecular and substrate states. The mixing of electronic states also explains the diffuse appearance of the TNAP orbitals in the STM images.

\subsection{Pb-TNAP coordination network on Pb(111)}

\begin{figure*}
\includegraphics[width=0.9\textwidth,clip=]{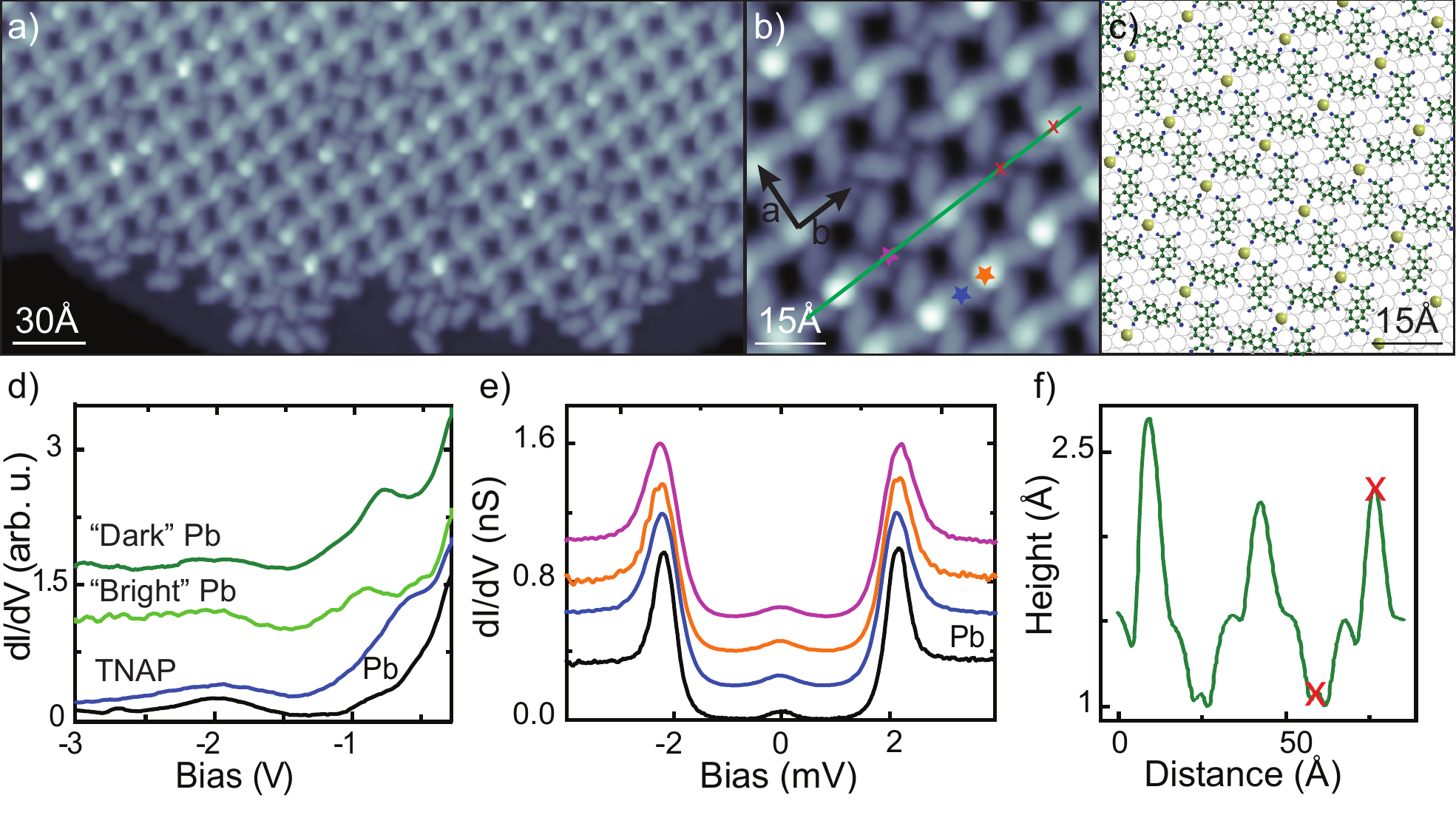}
\caption{a) Large-scale STM image of the porous Pb-TNAP network on Pb(111). b) Zoom into the Pb-TNAP structure with the lattice vectors indicated by arrows $a$ and $b$ ($I=0.5$~nA, $U=10$~mV). c) Structural model according to the STM images, which provides a commensurate structure with the Pb atoms sitting on hollow sites.  d) Constant-current $dI/dV$-spectra acquired over the different Pb atoms and TNAP molecule ($I=0.5$~nA, lock-in modulation $V_\mathrm{rms}= 5$~mV)~\cite{note2}. e) Constant-height $dI/dV$-spectra around Fermi level acquired over "dark" and "bright" Pb atoms and a TNAP molecule (feedback opened at $I=0.5$~nA, $U=10$~mV, $V_\mathrm{rms}=0.1$~mV). The location of the low-energy spectra are indicated by colored stars. The spectra in (d, e) are offset for clarity. f) Height profile along the line indicated in (b).} 
\label{fig2} 
\end{figure*}

We observe a distinctly different structure when the TNAP molecules are deposited at slightly higher temperature, i.e., at 330~K [Fig.~\ref{fig2}(a)]. Instead of the densely-packed molecular island, the resulting structure exhibits open pores. The TNAP molecules can be identified by their size and shape as flat lying in a windmill-shaped arrangement around fourfold nodes, where four cyano groups point towards each other [Fig.~\ref{fig2}(b,c)]. The nodes are filled with small protrusions, which exhibit two different apparent heights in the STM images, i.e., $2.3\pm0.4$~${\textrm{\AA}}$ and $1.1\pm0.2$~${\textrm{\AA}}$, respectively. 
Such a bonding pattern is reminiscent of metal-organic coordination motifs, in which a metal atom acts as the bonding "glue" between the electrophilic cyano end groups~\cite{TsengPRB09, TsengJPCC11, FaraggiJPCC12, AbdurakhmanovaNatChomm12, YangACSNano14}. At this growth temperature, atoms from the step edges possess sufficient thermal mobility for reacting with the TNAP molecules. The formation of similar metal-organic networks frequently occurs on noble metal substrates, where the $3d$ transition metal atoms from the substrate attach to electrophilic molecular endgroups and support a coordination bond~\cite{PerrySurfSci98, PawinAngew08, BjoerkPCCP10, XiaoJPCC12, FaraggiJPCC12, SirtlJacs13, UmbachPRB14, YangACSNano14, RodriguezJChemPhys15}.

The Pb-TNAP network is fully commensurate with the substrate. Its structure can be described by the lattice vectors $a=b=15~{\textrm{\AA}}$ enclosing an angle of $\alpha=85^{\circ}$. The Pb adatoms can thus sit in hollow sites, which would also be their energetically preferred adsorption site in an isolated case~\cite{LianPRB10}. The arrangement of the TNAP molecules is stabilized by their coordination to the Pb adatoms as well as by intermolecular electrostatic N--H bonds with the typical length of $2.2\pm0.2~{\textrm{\AA}}$. The N--N distance amounts to $5.6\pm0.2~{\textrm{\AA}}$, resulting in a minimum Pb--N bond length of $2.8\pm0.2~{\textrm{\AA}}$, assuming that the Pb atoms lie in the molecular plane. 
The two different regimes of apparent heights of the Pb adatoms suggest a buckling of the Pb atoms out of the molecular plane, thus leading to small differences in the three-dimensional bonding configuration. To investigate whether these are correlated with differences in the electronic structure, we recorded tunneling spectra on the Pb nodes [Fig.~\ref{fig2}(d)]. The unoccupied states do not show any deviation from the plain molecules (not shown). In the occupied states, we find a resonance at -0.9~eV on the "bright" Pb adatoms, which is slightly shifted towards the Fermi level on the "dark" species. These features are unique to the Pb-TNAP network; single Pb atoms on Pb(111) do not show such a resonance~\cite{LianPRB10}. The energetic shift agrees with a topographically lower "dark" atom, where the screening of the metal substrate lowers the ionization potential probed by the tunneling electrons~\cite{TorrenteJPCM08}. 

The two regimes of different Pb adatom height suggest two regimes of bonding configurations with rather broad bond length distributions. The regimes agree with the typical Pb--N bond lengths in three-dimensional cyano--Pb bonding motifs~\cite{KatzJACS05, GilPolyhedron12}. 
The large radius and $p$-block valence electron configuration of Pb allows it to adapt several bonding geometries and even different oxidation states. This is the reason for the wealth of Pb based three-dimensional metal-organic frameworks~\cite{ShimoniInorgChem98, YangChemEurJ07} and adaptability in surface-anchored Pb-organic networks~\cite{LyuJMCC15}. However, probably the closest resemblance of the coordination motif is borne to Pb-phthalocyanines, where the Pb core is ligated by four N atoms in a square pyramidal fashion~\cite{PapageorgiouPRB03}.

One possible bonding type would be ionic due to charge transfer from the Pb atom to the acceptor molecule TNAP. To check for a possible sign of this, we acquired tunneling spectra in the superconducting energy region. Again, we do not detect any sign of Shiba states, which could signify the presence of an unpaired electron spin and hence single charge occupation. Another possible bonding would be of more covalent character. The bonding angle between C--N--Pb being 180$^\circ$ is in agreement with a directional bonding pattern, similar to windmill structures of transition metal atoms and tetracyanoquinodimethane (TCNQ)-- the parent compound of TNAP~\cite{TsengJPCC11, AbdurakhmanovaNatChomm12}.

\subsection{Na-TNAP metal-organic network}

\begin{figure}
\includegraphics[width=8cm,clip=]{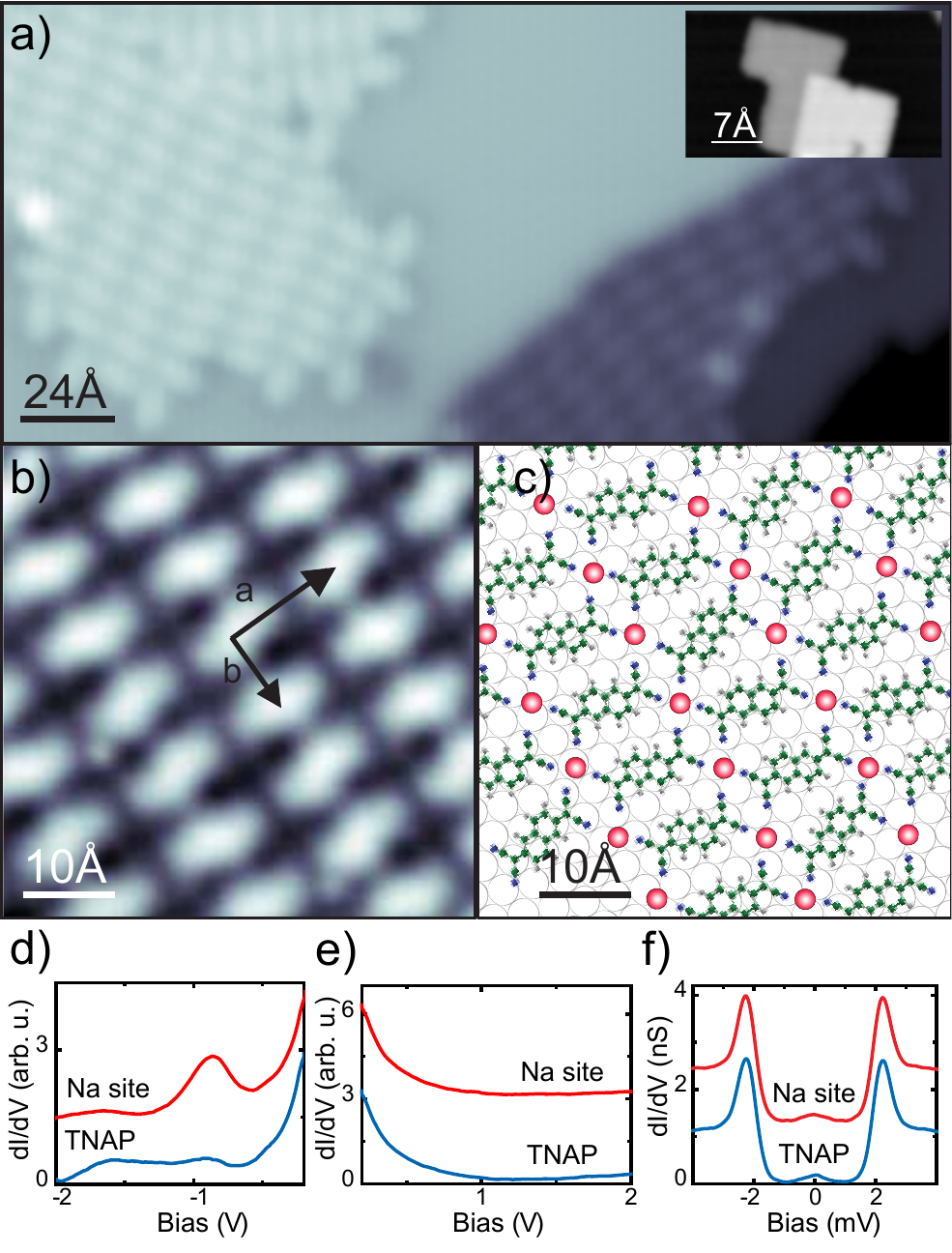}
\caption{a) STM image of the Na-TNAP islands on Pb(111) ($I=0.017$~nA, $U=1$~V). The inset shows an example of the previously prepared NaCl islands. b) STM zoom-in on the ordered Na-TNAP island ($U=-11$mV). c) Proposed structure of Na-TNAP on Pb(111). d, e) Constant-current $dI/dV$-spectra acquired over the Na atom and TNAP molecule ($I=0.5$~nA, $V_\mathrm{rms}=5$~mV). f) Constant-height $dI/dV$-spectra around Fermi level acquired over Na atom and TNAP molecule (feedback opened at $I=0.3$~nA, $U=10$~mV, $V_\mathrm{rms}=0.1$~mV). The spectra in (d-f) are offset for clarity.}
\label{fig3}
\end{figure} 

Motivated by the observation of ionic bonding between TCNQ and alkali metal atoms~\cite{AbdurakhmanovaNatChomm12, WaeckerlinChemComm2011, UmbachNJP13}, we explore this possibility for TNAP in the following.
As an alkali metal source, we deposit NaCl on the Pb(111) surface at room temperature. This leads to the formation of bi-layer islands of NaCl, similar to their growth on noble metal substrates [see inset of Fig.~\ref{fig3}(a)]~\cite{ReppPRL04, LauwaetJPCM12, HeidornNanoLett14}. Subsequent deposition of TNAP at room temperature results in a dissolution of the NaCl islands and the formation of a new TNAP network structure. 
 We find rectangular patterns of TNAP with lattice vectors $a=9\pm1~{\textrm{\AA}}$ and $b=14\pm1~{\textrm{\AA}}$ [Fig.~\ref{fig3}(a)]. These structures include fourfold nodes of cyano endgroups. These can only be stabilized by the inclusion of an adatom which mediates a bonding configuration and thus overcomes the electrostatic repulsion of the electron-rich cyano terminations. Because Pb atoms lead to a different bonding pattern (see above), the incorporated atom can be assigned to Na. The alkali atom is prone to form a charge transfer complex with the TNAP, while the chlorine desorbs in molecular form from the surface at room temperature~\cite{WaeckerlinChemComm2011}. The islands again consist of both enantiomers as reflected by the molecular shape in the STM images [Fig.~\ref{fig3}(b)]. Both orientations allow for the quinodimethane unit to align along the [110] crystal direction of the Pb substrate. This imprints the preferential orientation of the islands on the Pb(111) surface. However, we do not find any commensuration of the whole Na-TNAP layer with the substrate [Fig.~\ref{fig3}(c)].

The dissolution of the ionically bonded NaCl islands indicates that the preferentially formed Na-TNAP islands exhibit an even larger bonding energy. Such a large bonding energy typically involves a strong ionic contribution, which is expected to arise from the charge transfer of the $3s$ electron of Na to the electrophilic TNAP molecule~\cite{WaeckerlinChemComm2011, UmbachNJP13}. To probe possible signatures of such a charge redistribution, we record differential conductance spectra on the Na and TNAP sites. The spectra on the Na sites reveal a pronounced peak at -0.85~eV, whereas the TNAP spectra are essentially flat [Fig.~\ref{fig3}(d, e)]. If the $3s$ electron is donated from Na to TNAP, the density of states of the closed-shell cationic Na is expected to be depleted in a large energy range around the Fermi level~\cite{UmbachNJP13}. In contrast, the anionic TNAP species should exhibit its half-filled LUMO orbital close to the Fermi level. The unexpected resonance distribution can, however, be described within the charge transfer model by considering the three-dimensional extension of the molecular orbitals. The transferred electron is mainly localized at the electrophilic cyano groups. The electrostatic potential of the cationic Na atom leads to a deformation of the molecular orbitals and their extension over the Na sites. Thus, the corresponding molecular resonance appears on top of the Na. This scenario has been corroborated theoretically in the very similar system of Na-TCNQ~\cite{UmbachNJP13}. 

Differential conductance spectra around the superconducting energy gap on the Na sites do not exhibit Shiba states. This agrees with the closed-shell nature of Na after charge transfer [Fig.~\ref{fig3}(f)]. However, we also do not observe Shiba states on the TNAP. The reason for this absence may be manyfold: the hybridization of TNAP with the substrate may lead to a non-integer charge state or yield a potential scattering strength, where the Shiba states are not sufficiently far away from the superconducting gap edge to be detectable. Furthermore, the existence of Shiba states around an extended scattering potential due to an unpaired electron spin in an extended orbital has not been observed experimentally, neither discussed theoretically.

\subsection{Fe-TNAP metal-organic network}

\begin{figure*} 
\includegraphics[width=0.9\textwidth,clip=]{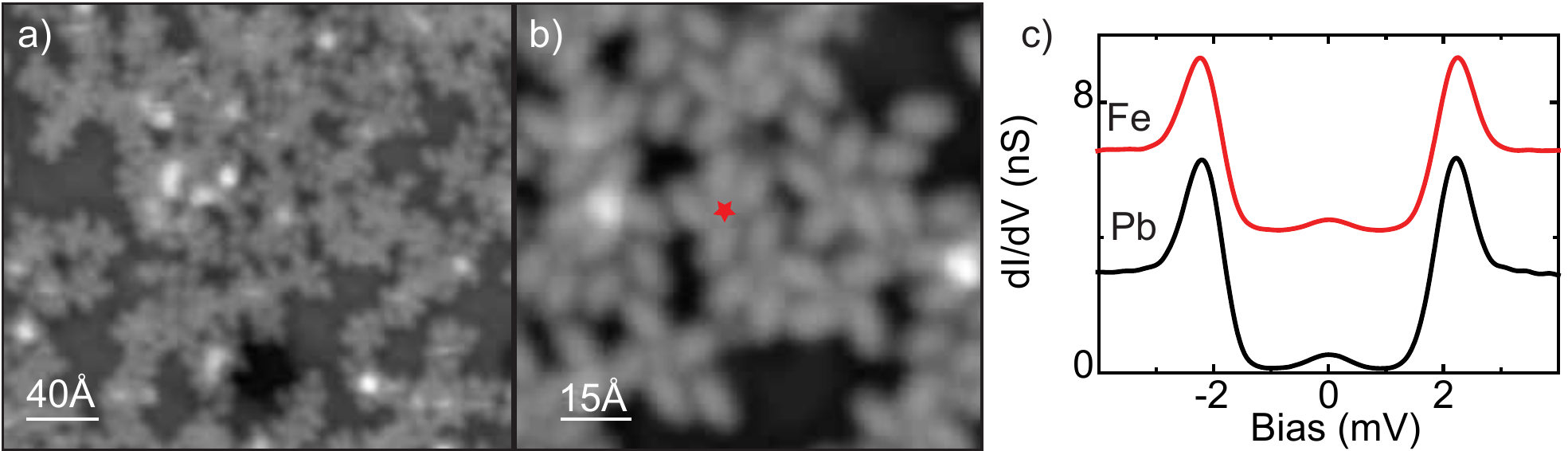}
\caption{STM images of the Fe-TNAP network after deposition at 160~K: a) overview ($I=0.21$~nA, $U=1$~V), b) zoom on the bonding nodes ($I=0.21$~nA, $U=100$~mV). c) Constant-height $dI/dV$-spectra acquired over the Fe indicated by a red star and Pb substrate (feedback opened at $I=0.21$~nA, $U=10$~mV, $V_\mathrm{rms}=0.1$~mV), spectra are offset for clarity.}
\label{fig4}
\end{figure*}

To avoid the complexity of an unpaired spin in a delocalized orbital, we next turn to the creation of a metal-organic network, in which a magnetic moment is expected to be carried by a transition metal atom. We co-deposit TNAP molecules and Fe atoms onto the Pb(111) substrate held at 160~K. The resulting adsorption structure is an irregular network as shown in Fig.~\ref{fig4}(a). We can identify different nodes, where cyano groups point towards each other, albeit with different molecular symmetry. These nodes reflect the incorporation of Fe atoms into the TNAP surrounding [Fig.~\ref{fig4}(b)]. 
We have tested a few of the bonding nodes for a magnetic fingerprint inside the superconducting energy gap. However, for those cases, we could not observe any hint of Shiba resonances [see for example Fig.~\ref{fig4}(c)].

\begin{figure*} 
\includegraphics[width=0.9\textwidth,clip=]{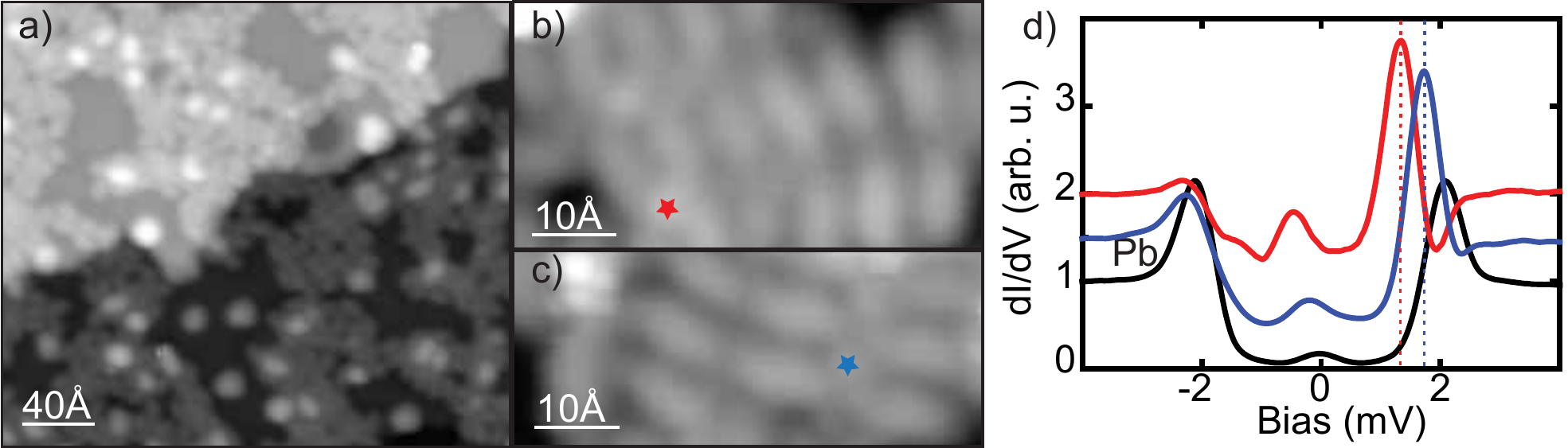}
\caption{STM images of the Fe-TNAP network after annealing to 340~K: a) overview ($I=0.21$~nA, $U=200$~mV), b,c) zoom on the bonding nodes [$I=0.1$~nA, $U=110$~mV for (b); $I=0.46$~nA, $U=24$~mV for (c)], d) Constant-height $dI/dV$-spectra acquired over Fe atoms in a molecular C2 bonding node (indicated by a red/blue star) and the Pb substrate. The spectra are normalized to the conductance outside the superconducting gap and offset for clarity. The Shiba states are found at a bias voltage $\Delta_\mathrm{tip}+E_b$ with the binding energies $E_b$, indicated by dashed lines.}
\label{fig5}
\end{figure*}

The high irregularity of the structure suggests insufficient mobility of the adsorbates for the creation of long range-ordered patterns. In an attempt to increase the order of the Fe-TNAP network we anneal the structures to 340~K. The overall appearance of the islands still remains highly irregular [Fig.~\ref{fig5}(a)]. Closer inspection of the bonding motifs reveals the occurrence of bonding nodes around which the TNAP molecules are arranged in a rectangular fashion [Fig.~\ref{fig5}(b,c)]. We suspect that these centers are Fe atoms, because Pb was shown to prefer a different coordination symmetry (see above). 
Differential conductance spectra on some of these Fe sites with local C2 symmetry show resonances inside the superconducting energy gap, i.e., at bias voltages $\left|V_{bias}\right|<(\Delta_\mathrm{tip}+\Delta_\mathrm{sample})/e$ [Fig.~\ref{fig5}(d)]. These always show up as a pair of resonances at energy $\pm(\Delta_\mathrm{tip}+\left|E_b\right|)$ with $E_b$ signifying the binding energy of the Shiba state, while the BCS-like resonances at $\pm(\Delta_\mathrm{tip}+\Delta_\mathrm{sample})$ disappear. At our measurement temperature of 4.5~K, quasiparticles can be thermally excited and therefore contribute to resonances at $\pm(\Delta_\mathrm{tip}-\left|E_b\right|)$~\cite{FrankeScience11, RubyPRL15a}. The intensity within the pairs is highly asymmetric reflecting the electron-like and hole-like weight of the wavefunction of the Shiba state in the weak tunneling regime~\cite{RubyPRL15a}. The position of the resonances, i.e., the binding energy reflects the strength of the magnetic scattering potential~\cite{FlattePRB97}. Indeed, we observe different binding energies on different Fe atoms [see e.g. Fig.~\ref{fig5}(d)]. Unfortunately, the high irregularity of the structure does not allow for a systematic investigation of the conditions for finding Shiba states and their binding energies. Their sole occurrence in the rectangular lattice sites suggests the necessity of a certain crystal field around the Fe atom. However, also the adsorption state of the neighboring TNAP molecules may affect the magnetic properties of the embedded Fe atom. Different adsorption sites on the underlying Pb lattice may further influence the magnetic coupling strength. Indeed, it has been observed that the different adsorption sites in a Moire structure formed by Manganese-phthalocyanine molecules on Pb(111) impact the magnetic exchange coupling strength~\cite{FrankeScience11, HatterNatComm15}.  

Unfortunately, the creation of long-range ordered structures of Fe-TNAP was not possible. This is probably inhibited by the incorporation of both Fe and Pb adatoms, which prefer different bonding symmetries. Studies of metal-organic networks on thin superconducting Au films circumvent this problem and present a promising avenue for the creation of Shiba lattices, which may even exhibit topological properties~\cite{LuArxiv15}.

\section{Conclusions}
The Pb substrate has for long time eludet from studies of self-assembling processes. With the promising perspective of creating magnetic structures with topological properties, superconductors will certainly gain importance as substrates for metal-organic networks~\cite{RoentynenPRL15}. We have explored viable routes to create such networks based on the organic electron acceptor TNAP. We observed a very large mobility of Pb adatoms, which are prone to form coordination bonds with cyano-based organic linker molecules. This mobility hampers favorable kinetics for the formation of extended networks, where the bonding energy is of similar magnitude. In particular, this is the case for transition metals, such as Fe. An ionic bond, which is energetically favored over the Pb coordination bond, circumvents this problem. The combination of an alkalimetal (Na) with the molecule acceptor TNAP has proven to from long-range ordered structures. Despite the charge transfer, these structures did not exhibit fingerprints of a magnetic interaction with the superconducting substrate. In contrast, we have found the desired magnetic interaction with the superconductor on some Fe sites within the Fe-TNAP network. A systematic analysis of the energy level alignment of the Shiba states inside the superconducting energy gap in dependence of their surrounding was not possible due to the high irregularity of the network. 
Nonetheless, we expect that a careful design of the molecular linkers will allow for the creation of magnetic metal-organic networks with long-range order. The quest for a magnetic pattern which creates topological states at the interface to the superconductor will surely enforce the efforts towards this goal.

\subsection{Acknowledgements}
GA acknowledges a research scholarship from the DAAD. We also acknowledge financial support by the ERC Consolidator Grant "NanoSpin".


\begin{thebibliography}{unsrt}
\bibitem{note1} The orientation of the Pb lattice was determined from atomic resolution images in a different preparation and from Ne subsurface inclusions, which result in hexagonal shaped apparent depression/protrusions in the STM images due to the formation of quantum well states~\cite{RubyPRL15}.
\bibitem{note2} When approaching the Fermi level the tunneling barrier width is significantly reduced in the constant-current mode. This leads to an increase in the transmittivity of the junction, which is reflected by an increase in $dI/dV$-signal close to the Fermi level. This is in contrast to the constant-height $dI/dV$-spectra in Fig.~\ref{fig1}(d). For comparison of the spectroscopy modes see, e.g., Reference~\cite{ZieglerPRB09}.
\end{thebibliography}
\end{document}